\documentclass[twocolumn,prl,aps,showpacs]{revtex4}
\usepackage{xspace,amsmath,amsfonts,amsthm,amssymb,amsbsy,graphicx,color}

\begin{document}

\title{Feedback Control Using Only Quantum Back-Action}
\author{Kurt Jacobs}
\affiliation{Department of Physics, University of Massachusetts at Boston, 100 Morrissey
Blvd, Boston, MA 02125, USA}

\begin{abstract}
The traditional approach to feedback control is to apply forces to a system by modifying the Hamiltonian. Here we show that quantum systems can be controlled without any Hamiltonian feedback, purely by exploiting the random quantum back-action of a continuous weak measurement.  We demonstrate that, quite remarkably, the quantum back-action of such an adaptive measurement is just as effective at controlling quantum systems as traditional feedback. 
\end{abstract}

\pacs{03.65.Yz, 87.19.lr, 02.30.Yy, 03.65.Ta}
\maketitle

In optical systems, continuous quantum measurements now have the precision to realize quantum feedback control~\cite{Smith02, Armen02, Higgins07}, and measurements in nano-electro-mechanical systems are rapidly approaching this regime~\cite{LaHaye04, Katz06, Houck07, Thompson08}. We expect the ability to perform real-time measurement and feedback in nano-systems is will open up a fertile field of applications for quantum control (see, e.g.~\cite{Hopkins03, Ralph04, Clerk08, Jacobs09}). 

Feedback control is traditionally realized by applying forces to a system (that is, modifying the Hamiltonian) in response to information obtained. This procedure is used even if the information is not realized in a measurement~\cite{Jacobs08b}, but is passed directly to a second quantum system for processing, as in ``coherent" feedback~\cite{wiseman94, Yanagisawa03a, Nurdin07}. This paradigm comes to us from classical feedback control. It has also been shown that measurements alone can be used to control a system by ``dragging" it using the ``quantum anti-Zeno" effect~\cite{DJJ, Pechen06, Shuang08}. However, for this to be effective the rate at which the measurements act must be much greater than the time-scale of the dynamics of the controlled system. Such a regime is unlikely to be practical for most applications of  continuous-time feedback control. 

Here we show that in quantum systems, one can implement feedback control without applying forces via a Hamiltonian, but merely by continually adjusting a continuous quantum measurement made on the system. The mechanism for this is not the anti-Zeno effect, but the (random) quantum back-action of the measurement. This mechanism is much more powerful than the anti-Zeno effect, since it induces dynamics on the time-scale of the measurement itself. 

Controlling quantum systems using measurement back-action is made possible by two things. The first is that a gradient in diffusion acts similarly, although not identically, to a deterministic force. The second is that quantum systems can be measured in different bases. As we will see below, it is this that allows us to exploit diffusion gradients to realize control. That a diffusion gradient induces deterministic motion can be seen from the following simple one-dimensional Fokker-Planck equation for the probability density of a variable $x$, 
\begin{equation}
    \frac{\partial P}{\partial t} =   - v \frac{\partial P}{\partial x}  
    + \frac{1}{2} \frac{\partial^2}{\partial x^2} \left[ D(x) P  \right] =  - \frac{\partial J}{\partial x}. 
\end{equation}
Here the first term describes deterministic motion of $x$ at the rate $\dot{x} = v$, and the second term describes diffusion at the rate $D(x)$. The quantity $J$ is the probability current, in general  a function of $x$ and time $t$. Calculating $J$ we find that 
\begin{equation}
    J(x) =  \left( v - \frac{1}{2}\frac{\partial D}{\partial x} \right) P - \frac{D(x)}{2} \frac{\partial P}{\partial x} . 
\end{equation}
So we see that the gradient of the diffusion rate, $\partial D/\partial x$, generates a term in the probability current equivalent to a negative deterministic velocity. 

\begin{figure}[t] 
\leavevmode\includegraphics[width=0.9\hsize]{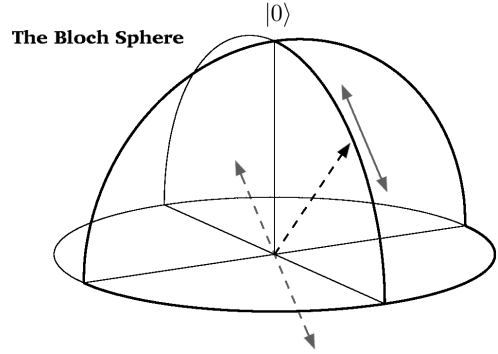}
\caption{A diagram depicting the control protocol. The target state is $|0\rangle$, at the top of the Bloch sphere. The black dashed line is the Bloch vector of the current state. The grey dashed line gives the direction of the measured spin, and the grey solid line gives the direction of the induced diffusion on the Bloch sphere.} 
\label{fig1} 
\end{figure}

We can realize a diffusion gradient by making a continuous measurement which is continuously modified using a feedback process (in other words, making an adaptive measurement). This technique is most easily understood by considering the control of a single qubit. To begin with, let us assume that the qubit is in an arbitrary state pure, and we wish to rotate it to a given ``target" state $|\psi\rangle$. We choose our coordinate axes so that $|\psi\rangle$ is the $\sigma_z$ spin-up eigenstate. We will call the spin-up state $|0\rangle$ and the orthogonal state $|1\rangle$.  An arbitrary pure state is $|\chi\rangle = \cos(\theta) |0\rangle + \sin(\theta)e^{i\phi}|1\rangle$, where $\delta = 2|\theta|$ gives the distance to the target state on the Bloch sphere. Since we have rotiational symmetry, we can set $\phi = 0$ without loss of generality, so that the state lies in the $xz$-plane: $|\theta\rangle \equiv |\chi\rangle  = \cos(\theta) |0\rangle + \sin(\theta)|1\rangle$. We now make a continuous measurement of spin, in the direction in the $xz$-plane that is perpendicular to the Bloch vector (see Fig.~\ref{fig1}). This generates the maximum rate of diffusion on the surface of the Bloch sphere, and keeps the state in the $xz$-plane. The equation of motion of the density matrix, $\rho_\theta = |\theta\rangle\langle\theta |$, under this measurement is~\cite{Brun02, JacobsSteck06} 
\begin{eqnarray}
   d\rho_\theta & = & -k(\theta)[\sigma_{\theta},[\sigma_{\theta},\rho_\theta] dt \nonumber \\ 
     &  & + \sqrt{2k(\theta)}( \sigma_{\theta}\rho_\theta + \rho_\theta\sigma_{\theta} - 2\langle \sigma_{\theta}\rangle \rho_\theta)dW ,  
\end{eqnarray}
where the spin component being measured is 
\begin{equation}
   \sigma_\theta =  \cos(2\theta) \sigma_x - \sin(2\theta)\sigma_z , 
\end{equation}
and $dW$ is the Wiener noise increment satisfying $dW^2 = dt$~\cite{JacobsSteck06}. Note that this measurement is continually determined by a feedback process, because the measured observable, and the measurement strength, $k$, depend on $\theta$; it is an {\em adaptive} measurement. With this measurement, the evolution equation for the distance to the target, $\delta$, is $d\delta = \sqrt{8k(\theta)}dW$. We can now obtain a diffusion gradient for $\delta$ simply by making the measurement strength a function of $\delta$. We choose $k = \kappa\delta^2$, and this results in the following linear stochastic equation for $\delta$: 
\begin{equation}
   d\delta = \sqrt{8\kappa} \delta dW . 
   \label{eq::delta1}
\end{equation}
The diffusion rate for $\delta$, as defined using the Fokker-Planck equation above, is now $D(\delta) = 8\kappa\delta^2$.  

\begin{figure}[t] 
\leavevmode\includegraphics[width=1\hsize]{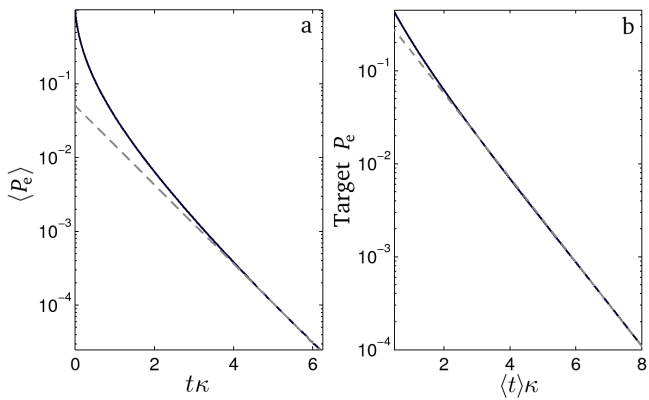}
\caption{Performance of quantum control with adaptive measurement according to two measures: (a) The average error probability, $\langle P_{\mbox{\scriptsize e}} \rangle$, as a function of time, when the system starts with $P_{\mbox{\scriptsize e}} = 1$. The average is taken over many noise realizations. The dashed line is the asymptotic slope. (b) The average time to reach a target value of $P_{\mbox{\scriptsize e}}$. The dashed line is the asymptotic slope.} 
\label{fig2} 
\end{figure}

The first questions we want to answer are 1. how fast this adaptive measurement moves the state to the target, and 2. how accurately it pins the state to the target once it has arrived. We consider two physically well-motivated ways to quantify the answers. The first is to calculate the evolution of the average probability that the system is not in the target state (the ``error" probability), $\langle P_{\mbox{\scriptsize e}}\rangle = \langle(1 - \cos(\delta))/2\rangle$. Here the average is taken over all realizations of the noise. When the system is close to the target state ($P_{\mbox{\scriptsize e}} \ll 1$), this becomes $\langle P_{\mbox{\scriptsize e}} \rangle \approx  \langle\delta^2 \rangle/4 = \langle \theta^2 \rangle $. To simulate the evolution we note that $\delta$ is a cyclic variable, so we must solve Eq.(\ref{eq::delta1}) on the interval $[0,\pi]$ with periodic boundary conditions. This has no analytic solution, so we solve it numerically. Choosing the initial state to be maximally distant from the target ($\delta(0) = \pi$), we plot the evolution of $\langle P_{\mbox{\scriptsize e}} \rangle$ in Fig.~\ref{fig2}. From this plot we see that after an initial transient, the average error probability decays asymptotically as a simple exponential. The rate of this decay is $1.23\kappa$. The adaptive measurement thus moves the system to the target on the timescale of the measurement, $1/k$. There is also no upper bound to the precision with which the measurement can pin the system to the target -- as time goes by $\langle \delta^2 \rangle$ and $P_{\mbox{\scriptsize e}}$ decrease without limit. 

\begin{figure}[t] 
\leavevmode\includegraphics[width=1\hsize]{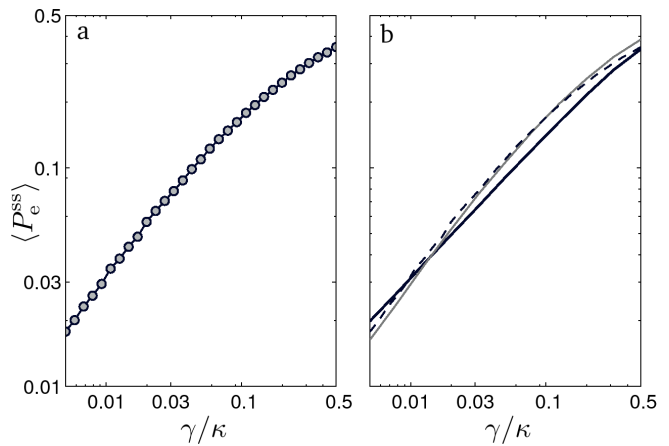}
\caption{The steady-state performance of three feedback protocols in the presence of dephasing noise and spontaneous decay. The performance is characterised by the average steady-state error probability, $\langle P_{\mbox{\scriptsize e}}^{\mbox{\scriptsize ss}} \rangle$. The noise strength is parameterized by $\gamma$, and the measurement and feedback strengths of all protocols are bounded by a constraint proportional to $\kappa$. (a) The diffusion-gradient protocol. (b) The diffusion-gradient protocol (dashed-line) along with two Hamiltonian feedback protocols. Dark line: Hamiltonian feedback with a perpendicular measurement; Gray line: Hamiltonian feedback with a parallel measurement. } 
\label{fig3} 
\end{figure}

A second well-motivated measure of the speed of the control protocol is the average {\em time} it takes for the control to bring the system within a given distance of the target state~\cite{Wiseman06x}. Once again, a sensible measure of this distance is the error probability $P_{\mbox{\scriptsize e}}$. In Fig.~\ref{fig3} we plot this average time as a function of $P_{\mbox{\scriptsize e}}$, once again starting the system furthest from the target. This shows very similar behavior to that of $\langle P_{\mbox{\scriptsize e}} \rangle$: the ``target" error probability, $P_{\mbox{\scriptsize e}}$, drops asymptotically as an exponential function of the average time to reach it. In this case the decay rate is $1.05\kappa$. 

So far we have evaluated the performance of our ``diffusion-gradient" control algorithm in the absence of noise. We now consider controlling a qubit in the presence of dephasing noise and spontaneous decay, these being the most common noise sources for this system. These noise sources are described by the master equation 
\begin{equation}
   \dot{\rho} = \!\!\! \sum_{j = x,y,z} \!\!\!\! \beta_i [\sigma_j,[\sigma_j,\rho]]  + \gamma ( 2 \sigma_- \rho \sigma_+  -  \{ \sigma_+ \sigma_- , \rho \} ) , 
\end{equation}
where $\sigma_\pm = (1/2)(\sigma_x \pm i\sigma_y)$, and $\{\cdot,\cdot\}$ is the anti-comutator. The parameters $\beta_i$ give the respective rates of dephasing in the three orthogonal directions, and $2\gamma$ is the rate of decay from the target state $|0\rangle$ to the state $|1\rangle$ which is orthogonal to it. The presence of noise mixes the state, reducing the length of the Bloch vector. Because of this our control protocol now needs to purify the system (lengthen the Bloch vector) to counter the effect of the noise, in addition to rotating the system to the target.  Fortunately these two tasks are compatible. 

We will denote the length of the Bloch vector by $a$, and use the quantity $\Delta \equiv 1 - a$ to quantify how mixed the state is. We note that the control protocol we have used so far already purifies the system, and in fact does so deterministically~\cite{rapidP}. For small $\Delta$, the evolution of the length of the Bloch vector is $\dot{\Delta} = - 8 k \Delta  = -8\kappa \delta^2 \Delta$. However, since $\delta \rightarrow 0$ as we approach the target state, we do need to add something else to our protocol to maintain the purity when the system is close to the target. We can do this without interfering with the dynamics of $\delta$ by adding a measurement {\em along} the direction of the Bloch vector. This measurement lengthens the Bloch vector by inducing a diffusion gradient for $\Delta$. The equations of motion for the system, when we add this additional measurement, and including the noise, are 
\begin{eqnarray} 
    d\delta & = & \sqrt{8\kappa} (\delta/a) dW \label{eq::delta2} \\ 
                &  & + \left\{ 2\gamma/a + C_\delta [ \gamma + 4(\beta_x - \beta_z) ] \right\} S_\delta dt , \nonumber \\ 
    da & = & (1 - a^2) \left[ 4 \kappa (\delta^2/a) dt + \sqrt{8\mu}\, dV \right]  \label{eq::Delta} \\ 
      &  & - 2 \gamma C_\delta dt - a ( \gamma + 4 \beta_x + 4 \beta_y ) dt \nonumber \\ 
      &  & - a (\gamma + 4 \beta_x - 4\beta_z) C_\delta dt , \nonumber 
\end{eqnarray}
where $C_\delta \equiv \cos(\delta)$, $S_\delta \equiv \sin(\delta)$, $\mu$ is the strength of the measurement parallel with the Bloch vector, and $dV$ is the Wiener noise of the second measurement. This is uncorrelated with $dW$, so that the noises satisfy the Ito calculus relation $dWdV=0$. 

We can expect this new protocol to act on the same time-scale as the previous protocol, since it exploits the same control mechanism for both $\delta$ and $\Delta$. What is of most interest now is the steady-state that the protocol achieves in the presence of the noise. We want to know whether we can achieve effective control, and if so, how strong the measurement strengths $\kappa$ and $\mu$ must be to accomplish this. Realizing good control in the steady-state means that the steady-state average error probability, $\langle P_{\mbox{\scriptsize e}}^{\mbox{\scriptsize ss}} \rangle$, is much less than unity. In this regime, we have $\langle P_{\mbox{\scriptsize e}}^{\mbox{\scriptsize ss}} \rangle  = \langle (1-a\cos(\delta_{\mbox{\scriptsize ss}} ))/2 \rangle \langle \approx \Delta_{\mbox{\scriptsize ss}}\rangle /2 + \langle \delta_{\mbox{\scriptsize ss}}^2 \rangle/4$.  Since we must solve the equations of motion for $\Delta$ and $\delta$ numerically, we gain some insight by examining the dynamical equations in the regime of ``good control"~\cite{Li09}. By ``good control" we mean the parameter regime in which the control protocol can maintains $\Delta \ll 1$ and $\delta \ll 1$.   

Expanding Eqs.(\ref{eq::delta2}) and (\ref{eq::Delta}) to leading-order in $\delta$ and $\Delta$ the dynamical equations become
\begin{eqnarray}
    d\delta & = & [3\gamma + 4(\beta_x - \beta_z )] \delta dt + \sqrt{8\kappa}\delta dW  \\  
    d\Delta & = & 4(\gamma + 2\beta_x + \beta_y - \beta_z) dt  + 2\sqrt{8 \mu}\Delta dV 
\end{eqnarray}
We see from these equations that the effect of noise is to add deterministic (drift) terms to the dynamics that act to increase both $\delta$ and $\Delta$. From our analysis above we know that in the absence of noise both variables tend to zero as $t\rightarrow\infty$. While the drift terms for the two variables are qualitatively different (that for $\Delta$ is constant, while that for $\delta$ is linear) we will find that both variables now reach a non-zero steady-state in the presence of noise. 

We simulate the above control protocol for a range of noise strengths. We note that for the purposes of simulation it is best to convert the variables $\Delta$ and $\delta$ to the Bloch vector elements $a_x$ and $a_z$, as this gives numerically stable equations. For simplicity we set all the four noise strengths to be equal ($\beta_x = \beta_y = \beta_z = \gamma$). We fix the measurement strength $\kappa$, set $\mu = \kappa$, and vary the noise strength from $0.005\kappa$ to $0.5\kappa$. The resulting steady-state average  error probability as a function of the noise strength is displayed in Figure~\ref{fig3}(a). This shows that the protocol realizes good control so long as the measurement strength is significantly larger than the noise. 

We now compare the performance of ``diffusion-gradient" feedback control with that of conventional feedback control, in which a Hamiltonian is used to continually move the system towards the target state~\cite{Jacobs08b}. In fact, the optimal protocol for controlling a qubit using Hamiltonian feedback is not presently known, so we will compare our new method with two such protocols. In both protocols we continually adjust the Hamiltonian so that it always rotates the system directly towards the target state, as this is likely to be optimal. The form of the Hamiltonian is $H(t) = \hbar \, \alpha(t) \sigma_{\mathbf{n}(t)}/2$. Here $\alpha$ is the rate at which the Hamiltonian rotates the system on the Bloch sphere (the ``feedback strength"), and $\sigma_{\mathbf{n}(t)}$ denotes a spin operator in the direction of a unit vector $\mathbf{n}(t)$. In the diffusion gradient protocol the measurement strength $\mu$ is fixed at $\kappa$, and $k$ is bounded by $k \leq \mbox{max}(\kappa \delta^2) = \pi^2\kappa$. To compare with Hamiltonian feedback, we must therefore also bound the feedback strength $\alpha$. In doing so we emphasize that comparing diffusion-gradient feedback with Hamiltonian feedback is an imprecise affair, since it is not clear what Hamiltonian ``strength" one should equate with the measurement strength $k$. Nevertheless, for the purposes of comparison we set $\mbox{max}(\alpha) = \mbox{max}(k) = \kappa\pi^2$, which seems reasonable.

We must now select the measurement strategies for our two Hamiltonian feedback protocols. For $\alpha \gg k$, and when the ratio $\alpha/k$ is fixed, previous work has shown that the optimal protocol is to measure in a basis perpendicular to the Bloch vector~\cite{Shabani08}. But we do not expect this to be optimal for the constraints we have here. In view of this we will examine two quite different measurement strategies. The first involves making a perpendicular measurement, for which we impose the bound $k \leq \kappa\pi^2$. It is important to note that for a given noise strength and fixed $\alpha$, this protocol will have an optimal value of $k$. We therefore perform simulations for a range of $k$ values, and report the best performance for each noise strength. The second strategy involves making a parallel measurement. Since the strength of the parallel measurement in our diffusion-gradient protocol was fixed at $\mu = \kappa$, we use the same measurement strength for this ``parallel" protocol.  

In Figure~\ref{fig3}(b) we plot the performance of the two Hamiltonian feedback protocols against that of the diffusion-gradient protocol. The performances of all of these are remarkably similar, to the surprise of the author. Thus according to the metrics we have used, adaptive measurement is just as effective for controlling systems as Hamiltonian feedback. 

Diffusion-gradient control can be extended easily to systems of any finite dimension $N$.   Consider an $N$-dimensional system that we wish to keep close to a target state $|\psi\rangle$. If the current state is $|\chi\rangle$, then we can apply a diffusion gradient to rotate $|\chi\rangle$ to $|\psi\rangle$ by measuring the appropriate observable, $\Sigma$. If the density matrix for the system is mixed, then we equate $|\chi\rangle$ with the eigenvector of the density matrix with the largest eigenvalue. We can calculate the required observable $\Sigma$ by considering the Bloch sphere of the two-dimesnional space spanned by $|\chi\rangle \equiv |0\rangle$ and $|\psi\rangle \equiv \cos(\theta) |0\rangle + \sin(\theta)e^{-i\phi} |1\rangle$; this gives $\Sigma = \mathbf{n}\cdot\boldsymbol{\sigma}$, with $\mathbf{n} = (\sin(2\theta),-\cos(2\theta),0)$. The purity of the state can be maintained without interfering with the diffusion gradient by measuring an observable that has $|\chi\rangle$ as one of its eigenvectors. This is an $N$-dimensional equivalent of measuring an observable parallel to the Bloch vector. 

The above prescription for controlling an $N$-level system does not fully specify the measurements --- there is much freedom left in the choice of observables. Further, in large systems, the choice of available observables will usually be restricted. How to optimize diffusion gradient control for larger systems, and under precisely what restrictions effective control can be achieved, are interesting questions for future work. 

{\em Acknowledgments:} This work was performed using the supercomputer in the College of Science and Mathematics at UMass Boston.  


\end{document}